**PAPER • OPEN ACCESS**

# On the gas heating effect of helium atmospheric pressure plasma jet

View the article online for updates and enhancements.

## You may also like

- Self-consistent time dependent vibrational and free electron kinetics for $CO_2$ dissociation and ionization in cold plasmas
  M Capitelli, G Colonna, G D'Ammando et al.

- Influence of $N_2$ on the $CO_2$ vibrational distribution function and dissociation yield in non-equilibrium plasmas
  L Terraz, T Silva, A Morillo-Candas et al.

- Discharge and optical characterizations of nanosecond pulse sliding dielectric barrier discharge plasma for volatile organic compound degradation
  Nan Jiang, Lianjie Guo, Kefeng Shang et al.

# Physica Scripta

PAPER

# On the gas heating effect of helium atmospheric pressure plasma jet

Fellype do Nascimento[1,2] 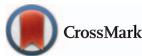, Torsten Gerling[2] and Konstantin Georgiev Kostov[1]

[1] Faculty of Engineering in Guaratinguetá, São Paulo State University-UNESP, Guaratinguetá 12516-410, Brazil
[2] ZIK plasmatis, Leibniz Institute for Plasma Science and Technology-INP, 17489 Greifswald, Germany

E-mail: fellype@gmail.com

## Abstract
Plasma medicine is an emerging research field which has been driven by the development of plasma sources suitable to generate low temperature plasmas. In many cases, such devices can operate without a gas flow, producing a plasma discharge from the ambient air. However, the most common case is the use of a working gas at a given flow rate to produce a plasma jet. Helium (He) is one of the gases commonly used as the carrier gas to generate cold atmospheric pressure plasma jets (CAPPJs) due mainly to the easiness to ignite a gas discharge with it. However, in this work it was observed that most of the heating of a He CAPPJ can come just from the expansion of the He gas. This was found through measurements of gas temperature ($T_{gas}$), using fiber optic temperature (FOT) sensors, and thermal output, using both FOT and infrared imaging with the He flow impinging on a copper (Cu) plate. Such findings were achieved through comparisons of $T_{gas}$ and the temperature on the Cu ($T_{Cu}$) plate in the conditions with and without discharge ignition, as well as comparing $T_{gas}$ in the free gas/jet mode with and without discharge ignition. It was verified that the $T_{gas}$ values increased as the distance from the gas outlet was enlarged, especially at low He flow rates, even without discharge ignition. Despite the temperature increase with distance, it is possible to produce plasma jets with temperatures lower than 40 °C at low He flow rates.

## 1. Introduction

Cold atmospheric pressure plasma jets (CAPPJs) have been successfully employed for treatment of a large sort of materials and in plasma medicine over the last two decades [1–6]. When applied on non-biological substrates the CAPPJs have both chemical and physical action on the treated surfaces by creating/removing functional groups and changing the surface roughness [7–10]. When interacting with biological tissues the plasma jets still have physical action, but the main interaction mechanism have been attributed to chemical reactions, like oxidative cellular stress, due to the generation of reactive oxygen and nitrogen species (RONS) within the plasma [4, 11, 12]. Some works also reported that pulsed electric fields generated within the CAPPJs have significant effects on biological targets, acting as a physical agent in synergy with other plasma properties and the RONS [13, 14]. Both physical and chemical actions of CAPPJs interacting with materials have significant dependence on the working gas [15, 16]. Air, argon (Ar) and helium (He) are the gases that are employed most often to generate CAPPJs, sometimes in admixtures with nitrogen ($N_2$), oxygen ($O_2$), water vapor or a different gas combination [11, 17]. The use of He as the carrier gas is one of the easiest ways to ignite a plasma discharge at atmospheric pressure using relatively low high voltage (HV) values. Only plasma jets generated with neon (Ne) usually have a lower ignition voltage than that of He for the same plasma device and operating conditions [18]. The choice of the working gas depends not only on the easiness in the plasma ignition, but also on factors like the gas temperature ($T_{gas}$) in the discharge phase, discharge power and current, production of reactive species and operation costs. In one of the first works comparing the use of He and Ne to produce plasma jets, it was found that the plasma propagation velocity can be higher when using Ne [19]. In addition, when considering endoscopic applications gas diffusion in tissues is also a critical factor to be considered, since He can diffuse in tissues, eventually in blood vessels or into the lungs differently from Ne, for example [20].

When a plasma jet is applied in the plasma medicine context, one of the most important parameters to be considered is the $T_{gas}$. Being that, for application in human tissues the $T_{gas}$ value should not exceed 40 °C in order to avoid cellular damage due to warming [21, 22]. Most works in the literature report that He is one of the working gases that produces atmospheric pressure plasma jets (APPJs) with $T_{gas}$ values close to the room temperature, that is, 300 K or 27 °C. Nonetheless, this is not always the case, since the $T_{gas}$ value in a plasma jet depends on several parameters like the discharge power and current, conductivity of the material on which the plasma impinges, gas flow rate (Q) and, eventually, on the room temperature [23, 24]. In addition, it is possible to cool down the working gas before using it to generate APPJs, which also can affect the gas temperature during the discharge phase. The employment of high Q values is one of the common ways to reduce the gas temperature [25, 26]. A shielding gas can also be used for such purpose [26].

Another important parameter to be considered when applying plasma jets for treatment of materials or tissues is the distance (d) between the plasma outlet and target. The d value influences many discharge parameters like temperature, discharge power, electrical current, electric field, as well as the production of reactive species, ultra violet (UV) light emission, etc [27]. Being that, all these parameters can contribute with the so-called plasma dose [28]. In fact, measurements of some plasma parameters as a function of d are required by the German DIN (German Institute of Standardization)-specification 91315. That specification can be seen as a complement of the IEC 60601-1 standard, as part of the criteria to certify plasma sources that are aimed for medical applications [29–31].

Gas temperature measurements in CAPPJs can be assessed through different techniques and equipment. The main difficult to measure this parameter is to obtain precise and accurate values without disturb or change other plasma jet properties. Optical emission spectroscopy (OES) is a non-disturbing technique that is used for $T_{gas}$ measurements. However, the maximum precision in this case is of the order of 10 K (10 °C), which is not good enough in some cases [32]. In some cases, infrared cameras (IR cams) can also be used as a tool to estimate $T_{gas}$ without disturbing the plasma. However, it is not so easy to capture the effluent image if $T_{gas}$ is close to the room temperature ($T_{room}$). The employment of thermocouples for $T_{gas}$ measurements are also reported on the literature [33–35]. Nonetheless, this kind of device usually has metallic components in the sensor probe. This can be a problem because plasma jets interact strongly with metals and, therefore, the results may not be accurate in this case. In addition, a thermocouple can also act as a floating electrode when in contact with the plasma jet, eventually influencing the discharge parameters [36, 37]. This can be avoided if the probe is covered with a dielectric material, as in [35]. However, in this case the response in time of the temperature readings became slower. Fiber optic temperature (FOT) sensors also have been used to measure $T_{gas}$ in plasma jets [27, 38–41]. Such kind of device does not have metallic compounds in the sensor region. Its tip is usually composed by a semiconductor and coated with a thin dielectric material. Thus, this will not present significant electrical interaction with the plasma. Nonetheless, such sensor can disturb the gas flow depending on the relative dimensions of both plasma jet and FOT. Besides that, FOT sensors can usually be assembled or chosen with appropriate dimensions and provide $T_{gas}$ measurements with an accuracy of the order of 0.2 °C when calibrated at room temperature.

It is known that helium and hydrogen ($H_2$) are the only gases that present negative values for the Joule-Thomson coefficient at room temperature, which makes a gas to heat up under adiabatic expansion at constant enthalpy [42–44]. It is also known that the He gas is subjected to buoyancy when it is flushed into the ambient air, mainly changing the propagation direction of the gas flow. This fact can affect some plasma jet parameters and the interaction of the electric field with the gas stream [45–48]. The buoyancy effect in He APPs tends to be less significant as the gas flow rate is increased [46]. Nevertheless, the effects related to these phenomena on the temperature of APPJs have not been fully taken into account in applications that employ He as the working gas, especially the medical and biomedical ones.

In this work, using FOT sensors to perform gas temperature and both FOT and infrared (IR) imaging for surface temperature and thermal output measurements. Through these measurements it was possible to identify that the main heating source of an APPJ that uses He as the working gas can be the He gas itself. It was found that the He gas temperature increases considerably as the distance from the gas outlet is incremented, with or without discharge ignition. This was observed in both free gas/jet mode and with a Cu plate in the He flow way. In the last case thermal energy was transferred to the target even without plasma discharge, that is, the Cu plate was heated up through simple exposure to the He flow. It was also found that the increment in $T_{gas}$ values as a function of d depends significantly on the gas flow rate.

## 2. Materials and methods

Gas temperature ($T_{gas}$) measurements were performed using helium (He, 99.999% purity) with ignition of a plasma discharge and also without that, that is, only the neutral gas temperature was measured in the last case. In both situations measurements were carried out in two ways: with the gas/plasma impinging on a thin copper (Cu) plate and in the free expansion/free jet mode. For the first case, the Cu plate temperature ($T_{Cu}$) was measured together with $T_{gas}$. For the free mode, the room temperature ($T_{room}$) was measured at the same time as $T_{gas}$. The Cu plate employed in the experiments is a square with sides measuring 10 mm and thickness of 0.3 mm, with a mass of 0.312 g. For comparison of the gas temperature behavior as a function of the distance from the plasma outlet, some additional $T_{gas}$ measurements were performed using argon (Ar) and carbon dioxide ($CO_2$), both 99.999% pure. In the case of Ar, the $T_{gas}$ measurement was performed only with discharge ignition, that is, producing a plasma jet. A mass flow controller (MKS, model 1179B) connected to a multi gas controller (MKS, model 647C) was used to control and adjust the gas flow rate. Figure 1(a) shows an overview of the plasma source, with the scheme for measurements of electrical parameters when producing a plasma jet and figure 1(b) presents a typical voltage waveform produced by the power supply. Figure 2 presents the scheme for the measurements of gas temperature. The plasma source shown in figure 1 consists of a dielectric barrier discharge (DBD) reactor to which a long and flexible plastic tube, with circular cross-section area, is connected. The DBD reactor is composed by a dielectric enclosure (with inner diameter of 10 mm) containing a pin-electrode (1.8 mm in diameter) made of tungsten encapsulated by a closed-end quartz tube with outer and inner diameters equal to 4.0 mm and 2.0 mm, respectively. The pin-electrode is in turn connected to a male metallic socket, which is attached to the dielectric enclosure and is plugged into the female socket of the power supply. Inside the long tube there is a thin copper wire (0.5 mm in diameter), which is fixed to a metallic connector placed inside the reactor chamber. This is a device that works using the jet transfer technique [49].

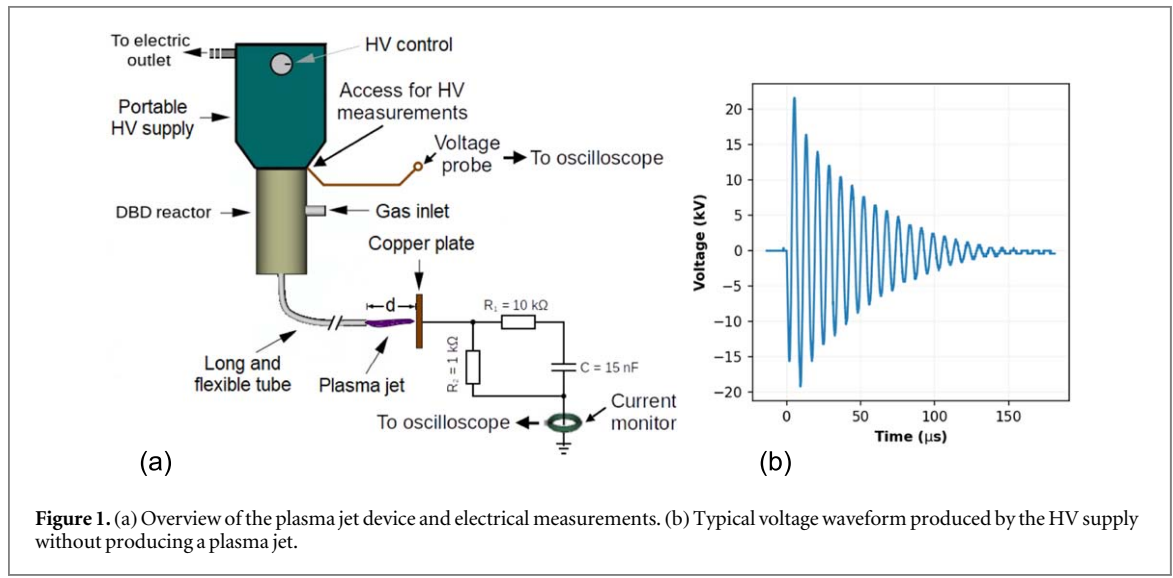

**Figure 1.** (a) Overview of the plasma jet device and electrical measurements. (b) Typical voltage waveform produced by the HV supply without producing a plasma jet.

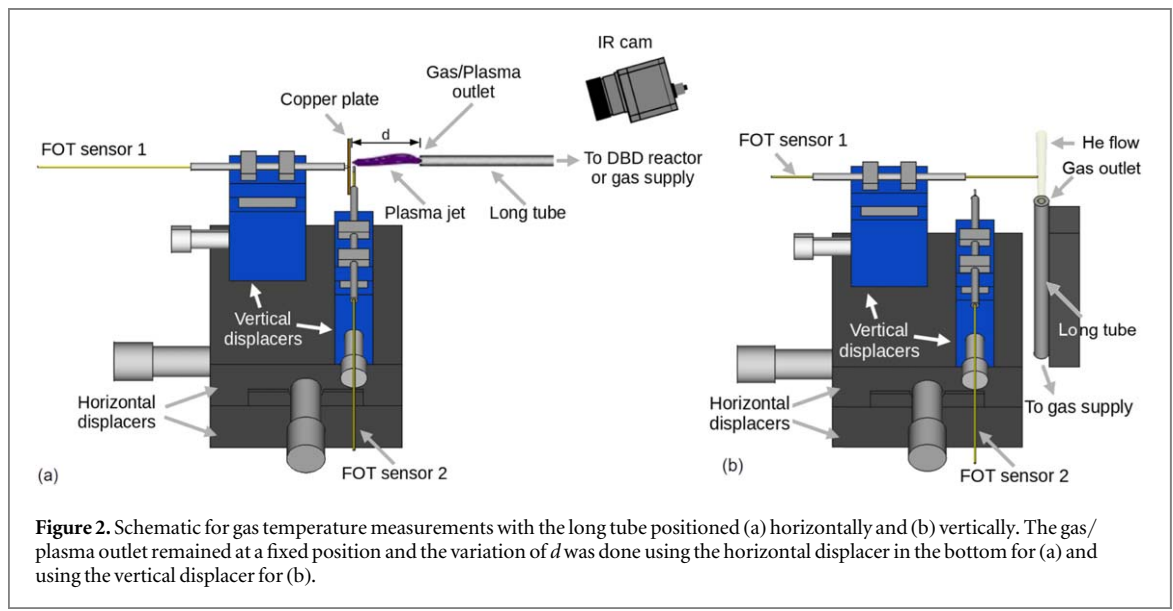

**Figure 2.** Schematic for gas temperature measurements with the long tube positioned (a) horizontally and (b) vertically. The gas/plasma outlet remained at a fixed position and the variation of *d* was done using the horizontal displacer in the bottom for (a) and using the vertical displacer for (b).

When the dielectric chamber is fed with a working gas (He in this work) and the high voltage (HV) is switched on, a primary plasma discharge is ignited inside the reactor. So, the plasma impinges on the metallic connector, which is not in touch with the quartz tube, and it acts as a floating electrode. Thus, the He gas flows through the long plastic tube and at its distal end a secondary discharge is ignited, producing a plasma jet.

In figure 2 it is depicted the scheme for placement of the fiber optic temperature (FOT) sensors, from LumaSense Technologies Inc. GmbH, USA (model FOT Lab Kit), employed for temperature measurements with the long tube oriented horizontally (a) and vertically (b). The data acquisition for gas temperature measurements were carried out as a function of the distance *d* from the gas/plasma outlet and also as a function of the gas flow rate (*Q*). When measuring as a function of *d*, the long tube remained at a fixed position and the *d* values were changed using the horizontal displacer at the bottom of figure 2 when the tube was oriented horizontally, or the vertical displacer otherwise. When measuring temperatures as a function *Q*, the *d* values remained fixed while the *Q* values were varied. The FOT sensors employed in this work are capable to measure temperatures with a precision of ±0.23 °C. Such FOTs can also be used to record the gas temperature values as a function of time, which allows us to calculate the thermal output ($P_{th}$) from the $T_{Cu}$ curve as a function of time by applying [31, 40]:

$$P_{th} = m \cdot c_p \cdot \frac{dT_{Cu}}{dt} \qquad (1)$$

where $m = 0.312$ g is the mass of the Cu plate and $c_p = 0.385 \mathrm{J \cdot K^{-1} \cdot g^{-1}}$ is the specific heat capacity of the copper. The time derivative $dT_{Cu}/dt$ is calculated in the time interval between plasma on and off, when a discharge is ignited, or between the time instants when the He gas flow is switched from off to on when measuring $T_{Cu}$ without discharge ignition. In part of the experiments, $T_{Cu}$ and $T_{gas}$, or $T_{Cu}$ and $T_{room}$, were recorded simultaneously using the FOT number 1 and 2, respectively. When measuring $T_{Cu}$ and $T_{gas}$ the distance *x* between the Cu plate and the FOT 2 was 1.0 mm.

In addition to the temperature measurements performed with the FOT sensors, an infrared camera (IR cam) from Optris GmbH (model optris® PI 450) was used to check the temperature and thermal output of the He gas impinging on the Cu plate surface. This equipment has a nominal accuracy of ±2.0 °C at 23 °C. In any case, the $T_{gas}$ and $T_{Cu}$ were measured as a function of the distance (*d*) from the gas/plasma outlet to the target. When the free gas expansion/plasma jet were under study, the $T_{gas}$ and $T_{room}$ values were recorded at the same axial distance (*d*) from the gas/plasma outlet. For the operated FOT sensors, data acquisition was performed with a home made Python software, PlaDinSpec (made with Python 3.8, Python Software Foundation, Wilmington, DE, USA), measuring and averaging 100 values.

The plasma source employed in this experiment uses a HV supply which is able to produce damped-sine waveforms with peak voltage values up to 20 kV and oscillating frequency of ∼110 kHz in the sinusoidal phase. This kind of waveform behaves as a pulsed HV signal with effective pulse duration (voltage higher than 5 kV) of nearly 70 μs (see 1(b)). Although the voltage oscillation inside each pulse is relatively high, the power supply is able to produce only three pulses with ∼1.7 ms separation between each two at a repetition rate of 50-60 Hz, depending on the electrical network frequency. A positive point regarding the low repetition rate of the HV pulses is that this prevents the excessive heating of the APPJ, which usually happens when high-frequency power sources are employed to produce plasma jets [50, 51]. This was important in this work to allow the observation of the temperature variation of the working gas as a function of the distance from the gas outlet. The electronic circuit shown in figure 1, formed by a 10 kΩ resistor in series with a 15 nF capacitor, with both connected in parallel to a 1 kΩ resistor, is aimed to simulate the electrical impedance of the human body. It is based on the IEC 60601-1 standards [29, 31, 40].

When producing a plasma jet, the voltage (*V*(*t*)) and current (*i*(*t*)) waveforms were recorded using a digital oscilloscope from Tektronix (model DPO 4104). The voltage was measured using a 1000:1 voltage probe (Tektronix, model P6015A) and the current was measured with a current monitor (Tektronix, model CT-1). The record length of the *V*(*t*) and *i*(*t*) waveforms was set to $10^5$ points, which allows measurements in relatively large time scales (4 ms) with temporal resolution of 40 ns. Discharge power ($P_{dis}$) values were calculated using the *V*(*t*) and *i*(*t*) data by applying:

$$P_{dis} = f \int_{t_0}^{t_f} v(t) \cdot i(t) \, dt \qquad (2)$$

where *f* = 50 Hz in our experiments. The effective current in the discharge was also calculated from the *i*(*t*) signals as:

$$i_{RMS} = \sqrt{\frac{1}{T} \int_0^T i^2(t) \, dt} \qquad (3)$$

with $T = 1/f$.

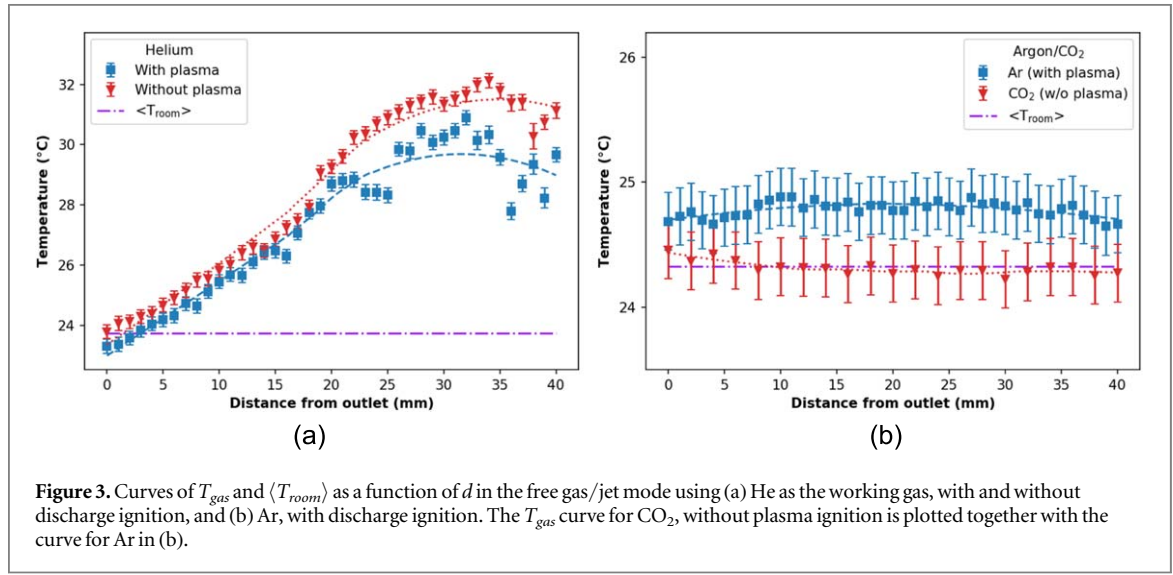

**Figure 3.** Curves of $T_{gas}$ and $\langle T_{room}\rangle$ as a function of $d$ in the free gas/jet mode using (a) He as the working gas, with and without discharge ignition, and (b) Ar, with discharge ignition. The $T_{gas}$ curve for $CO_2$, without plasma ignition is plotted together with the curve for Ar in (b).

The $V(t)$ and $i(t)$ waveforms were recorded for ten consecutive plasma discharges at different distances between the plasma outlet and the Cu plate and equations (2) and (3) were applied each time. Thus, the final discharge power and effective current values at each position are the average values of ten consecutive measurements. The Cu plate was electrically connected to the ground through the RC circuit shown in figure 1 in all the experiments, except when it is mentioned that it is directly grounded.

## 3. Results and discussion

In this section the results of the temperature measurements as well as the results from electrical characterization are presented. The results related to temperature measurements are separated in two situations: one of them with the gas or plasma jet not impinging on the Cu target (the free gas/jet mode) and the other with the gas/plasma directed to the Cu plate (conductive mode, when generating plasma).

### 3.1. Gas temperature without target (free gas/jet mode)

In figure 3 are presented results of gas temperature measurements as a function of the distance from the gas outlet in the free gas/jet mode. Figure 3(a) shows the curves obtained using He as the working gas, comparing the cases with and without discharge ignition and figure 3(b) shows curves of temperatures for Ar with discharge ignition and for $CO_2$ without plasma ignition. Such gas temperature measurements performed employing Ar and $CO_2$ are intended to obtain the behavior of $T_{gas}$ as a function of $d$ for different working gases and to compare it with the behavior of the gas temperature observed using He. The average room temperature ($\langle T_{room}\rangle$) is also indicated in the graphs of figure 3.

The measurements presented in figure 3 were performed with the gas tube in horizontal orientation. In the case when He was the working gas and with discharge ignition, we tried to position the FOT sensor so that it was as close as possible to the visual center of the plasma column. This was possible with the help of the light emitted by the plasma jet up to $d \approx 25.0$ mm, which is the approximate length of the plasma plume in the free jet mode. With He and without discharge ignition, we tried to place the FOT sensor in the center of the gas column with help of the visual contrast between the He flow and the ambient air due to the smaller refractive index of the first. When using Ar as the working gas, the FOT sensor was positioned in the same way as for He while for $CO_2$ the FOT was always in the same vertical position.

As it can be seen in figure 3(a), the trends of the $T_{gas}$ curves with and without discharge ignition are almost the same. There are some differences in the $T_{gas}$ values for larger $d$. However, such differences are mainly due to the difficulty to keep the FOT sensor always in the center of the gas column. A possible way to be sure that the FOT sensor is always in the center of the gas flow would be to combine temperature measurements using FOT sensors with an imaging technique like the Schlieren one. In figure 3(a) we also see fluctuations in the temperature values for $d$ larger than 20.0 mm in both curves. Such effect is possibly related to transitions of the gas stream from laminar to intermediate or turbulent regimes. However, this can also be due to the fact that FOT sensor is not anymore in the pure He flow, but mixed with the ambient air. Thus, in the context of the measurements, those differences in the $T_{gas}$ values with and without plasma are not significant. Nevertheless, it is possible that the lower temperature values obtained with plasma for $d$ higher than 22-25 mm have some influence from the gas flow modulation that can occur when the plasma is on [52, 53]. The small differences between the temperatures measured with and without plasma may also indicate that the electrical energy used to ignite the APPJ in free mode is not fully converted into thermal energy when producing the plasma plume.

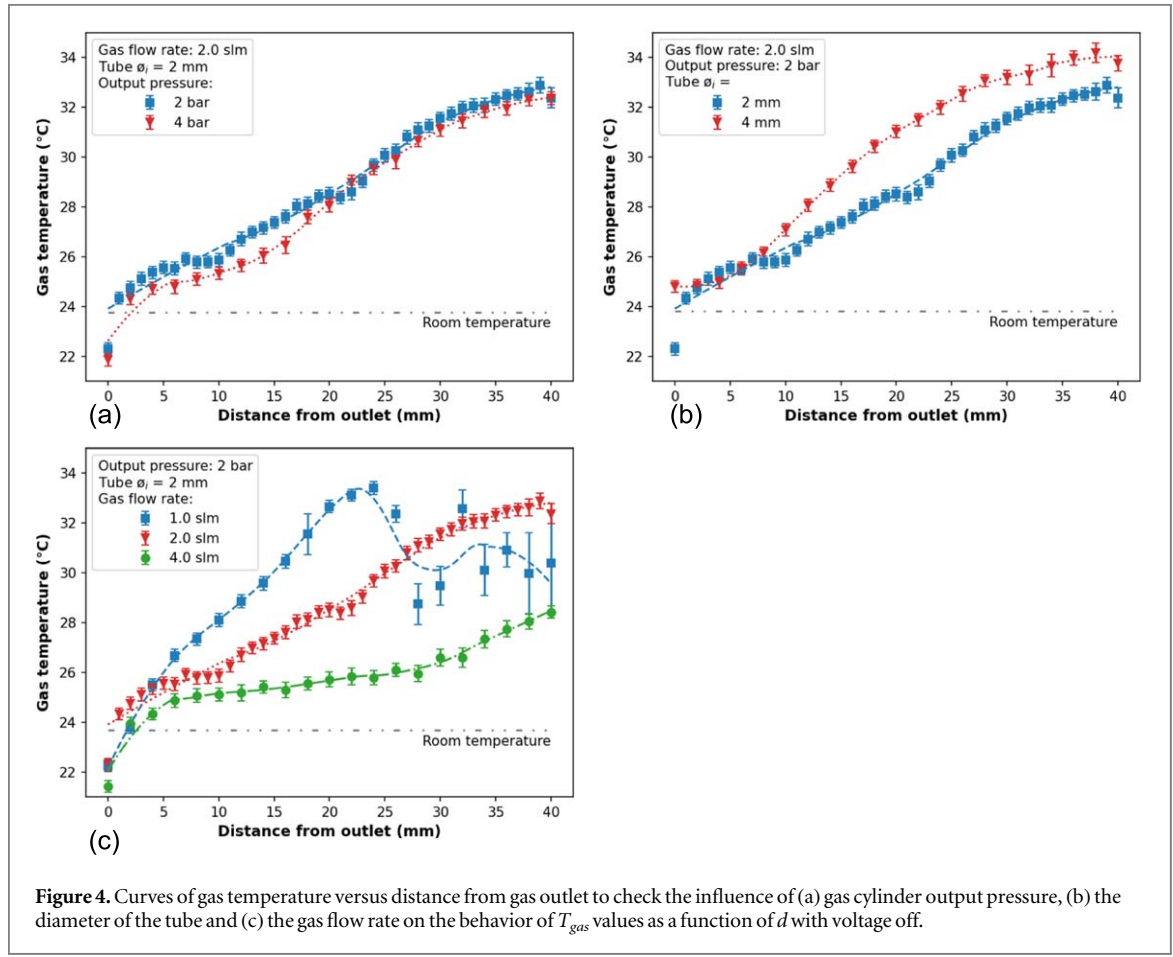

**Figure 4.** Curves of gas temperature versus distance from gas outlet to check the influence of (a) gas cylinder output pressure, (b) the diameter of the tube and (c) the gas flow rate on the behavior of $T_{gas}$ values as a function of $d$ with voltage off.

From figure 3(b) one can notice that the gas/plasma heating as $d$ increases does not happen for the investigated device when Ar is the working gas and there is plasma ignition, or when employing $CO_2$ without plasma discharge. With Ar as the working gas, we also notice that the $T_{gas}$ values in the free jet mode are very close to the $\langle T_{room} \rangle$ ones, while the temperature of the $CO_2$ gas is the same as $\langle T_{room} \rangle$. These results obtained using Ar and $CO_2$ can be understood as a first indicative that the gas temperature increments observed for He are not related to a problem with the FOT sensor.

The variation of the $T_{gas}$ values as a function of $d$ was also evaluated in a way to compare the possible influence of different parameters on $T_{gas}$. Figure 4 shows the results obtained for curves of $T_{gas}$ vs $d$ for the He gas flowing freely from the tube outlet without discharge ignition. In 4(a) different output pressure values were set in the gas cylinder (2 bar and 4 bar, for instance), in 4(b) plastic tubes with different inner diameters ($\phi_i$ equal to 2.0 mm and 4.0 mm) were employed and in 4(c) different gas flow rate values were applied ($Q$ equal to 1.0, 2.0 and 4.0 standard liters per minute (slm)). Being that when one parameter is changed all the others are kept constant. Under any of the stated conditions, it is observed that there is a significant variation of the $T_{gas}$ values as a function of $d$, with the curves growing monotonically in most cases in the measured distance range.

From figure 4(a) it can be seen that the $T_{gas}$ values increase almost linearly with $d$ despite the fact that the voltage is switched off, so without plasma ignition. Also, the curves for both output pressure of the gas cylinder do not differ significantly, which suggests that this parameter has no strong influence in the $T_{gas}$ values under the conditions in which the experiments were carried out. In figure 4(b) it can be noticed that the $T_{gas}$ values tend to be higher when the plastic tube has a large diameter (4.0 mm in this case). This is something interesting because the velocity ($v$) of the gas elements is inversely proportional to the tube diameter square at a constant gas flow rate:

$$v \propto \frac{Q}{A} = \frac{Q}{\pi \phi_i^2 / 4} \tag{4}$$

With this taken into account, we can infer that the lower the velocity of the He atoms when leaving the gas outlet, the higher the $T_{gas}$ values. This statement is partially supported by the results shown in figure 4(c), in which it is clearly seen that the lower the gas flow rate, resulting in lower $v$, the higher the $T_{gas}$. However, when comparing the supposed change in $v$ due to the change in $\phi_i$ or $Q$, the proportionality is not the same. This suggests that the particle number rate ($\eta$ = number of particles per unity time) leaving the outlet has stronger influence in $T_{gas}$, with lower $\eta$ leading to higher temperature values. Figure 5 shows the variation of the gas temperature as a function of $Q$ for some selected $d$ values. That measurements were performed changing the $Q$ values with $d$ fixed and generating datasets that are independent from the previous ones. In that figure it can be clearly seen that the $T_{gas}$ values tend to decrease as $Q$ is incremented. It is interesting to notice that for $d = 0.0$ mm the $T_{gas}$ values are lower than the $T_{room}$ ones for $Q > 1.0$ slm. In addition, for all measured distances at larger $Q$ the gas temperature tends to decrease to values close to or lower than $T_{room}$. These results enforce the evidence that both low velocity and low atom number rate leaving the gas outlet lead to higher $T_{gas}$ values.

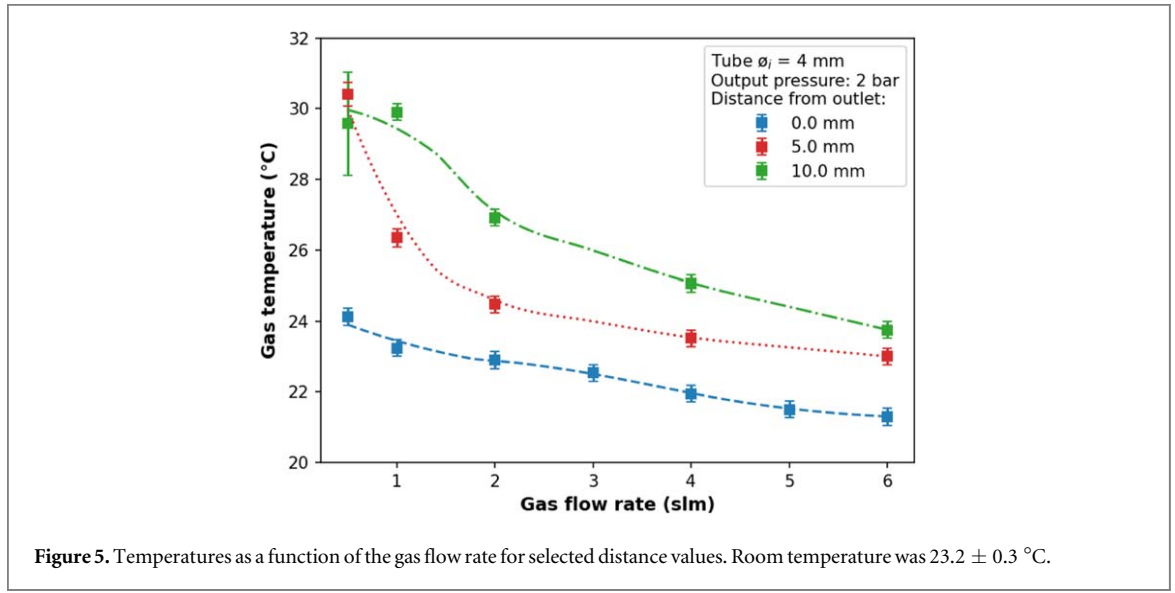

**Figure 5.** Temperatures as a function of the gas flow rate for selected distance values. Room temperature was 23.2 ± 0.3 °C.

In addition to the measurements shown in figures 4 and 5, we also checked the influence of the distance from the gas cylinder to the plasma outlet ($D_{CO}$) in the $T_{gas}$ variation with $d$. In this case, no significant differences were found when the $D_{CO}$ value was 4.0 m or 25.0 m.

It is important to note that the gas temperature measurements presented in figures 4 and 5 were performed with the tubes placed in a vertical orientation, with the outlet pointing upwards. The purpose of this was to try to ensure that the gas would be flowing in an approximately straight direction, which is not possible with the tube oriented horizontally when He is the working gas. In this case, because the He atoms are considerably lighter than the atoms and molecules that form the ambient air the trajectory of the gas column is shifted upwards when away from the gas outlet [46, 52].

It is well known that when the He gas is under adiabatic expansion at constant enthalpy it is subjected to the Joule-Thomson effect. In this case, the Joule-Thomson coefficient for He at room temperature presents negative values [42–44]. Due to that, it is expected a temperature increase when He is flushed from a cylinder into the ambient air. This may be one of the reasons for the results observed in figures 4 and 5, especially when $Q$ is ⩽ 1.0 slm, since at small $Q$ values the experimental conditions tend to be closer to one described in the Joule-Thomson experiment (a small amount of gas passing through a throttle). However, the Joule-Thomson effect does not fully explain the increment in the $T_{gas}$ values as $d$ is incremented.

Once the He atomic mass is much smaller than the mass of the environment air, when it is flushed into the ambient the buoyancy force accelerate the He atoms. This can be an additional factor for the temperature increase, which is partially supported by the fact that lower $Q$ values lead to higher $T_{gas}$. As the He mass being introduced into the ambient air is lower at low $Q$, the resulting acceleration of the outgoing mass becomes higher. Therefore, for large $d$ values the velocity of the He atoms is much higher than for small $d$ and, consequently, the temperature associated with $v$ is also higher. Although this is intuitive for the experiments performed with He being flushed in the vertical direction, when operating with the tube oriented horizontally the buoyancy effect is probably not significant. Another phenomenon that can possibly be affecting the $T_{gas}$ values when He is flushed into the ambient air is the diffusion thermoeffect, also known as Dufour effect, which causes a difference in temperature due to a gradient of species concentration [54–56]. Such gradient of species concentration is known to occur when gases are flushed into the air. Thus, the total gas heating may be due to a synergy between the buoyancy, Joule-Thomson and Dufour effects.

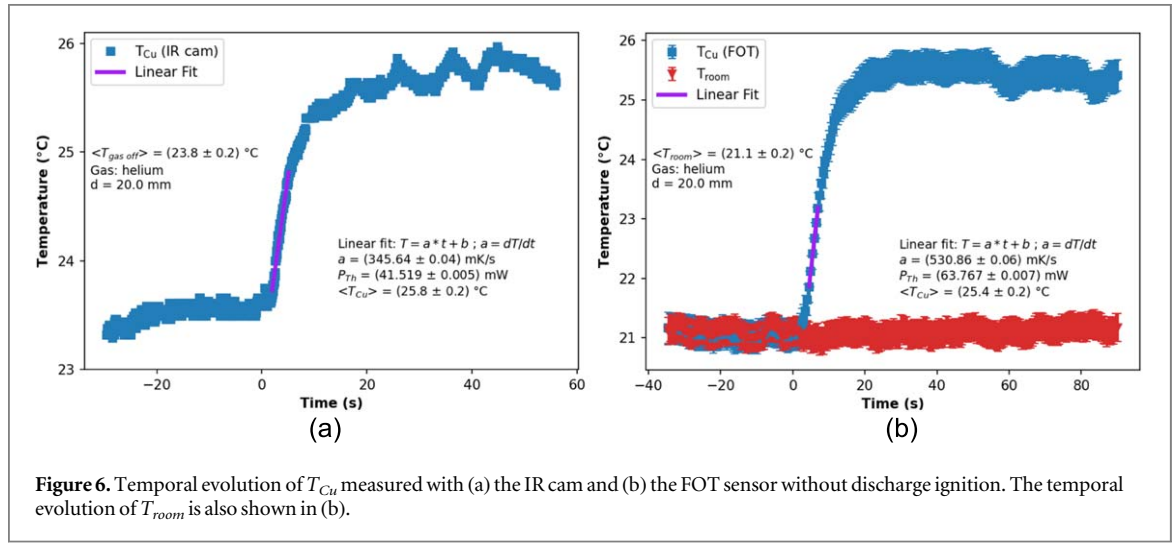

**Figure 6.** Temporal evolution of $T_{Cu}$ measured with (a) the IR cam and (b) the FOT sensor without discharge ignition. The temporal evolution of $T_{room}$ is also shown in (b).

### 3.2. Gas temperature with target and thermal output

Temperature measurements were also carried out with the gas impinging on a thin Cu plate using both the FOT sensor and an IR cam without discharge ignition. This was done in order to check if the gas temperature values observed in the previous section are related with a possible limitation on the use of FOT sensors when He is used as the working gas. The $T_{Cu}$ values were recorded as a function of time ($t$) in order to be possible to calculate the thermal output intrinsic to the gas ($P_{th-gas}$) from the inclination of the $T_{Cu}$ vs time curve right after the gas flow rate is turned on. Examples of curves showing the temporal evolution of the $T_{Cu}$ values, for $d = 20.0$ mm, are shown in figure 6 for (a) data measured using the IR cam and (b) data obtained with the FOT sensor. The behavior of the $T_{room}$ as a function of time using the FOT was also recorded and it is shown in figure 6(b). The gas flow rate employed on these experiments was 2.0 slm and the internal tube diameter was 2.0 mm.

From the $T_{Cu}$ versus time curves shown in figure 6 it can be seen that the He gas actually transfers heat to the Cu plate when the gas flow is switched on. Besides that, from the observation of heat transfer from the He gas to the Cu plate using two different measurement tools it is possible to discard the thesis that the FOT sensors used in this work are not suitable for gas temperature measurements when He is the working gas employed to produce APPJs.

Measurements of $T_{Cu}$ versus $t$ were carried out for different distances between gas outlet and target. Such data were also used to extract the average values of $T_{Cu}$ before and after switching on the gas ($\langle T_{gas\ off}\rangle$ and $\langle T_{Cu}\rangle$, respectively). Such averages were calculated in the plateau of the $T_{Cu}$ vs $t$ curves in the gas off (for $\langle T_{gas\ off}\rangle$) or gas on (for $\langle T_{Cu}\rangle$) phases. With those values it was possible to plot curves of $\langle T_{gas\ off}\rangle$ and $\langle T_{Cu}\rangle$, as well as curves of $P_{th-gas}$, as a function of $d$. The results are shown in figure 7 with (a) the behavior of $T_{gas}$ values and (b) the behavior of $P_{th-gas}$ as a function of $d$, comparing the measurements performed using the IR cam and the FOT sensor without discharge ignition. In our experiment it was verified that for the measurements performed with the FOT sensor the $\langle T_{gas\ off}\rangle$ values were approximately the same as the average $T_{room}$ ones ($\langle T_{room}\rangle$). From figure 7 it can be seen that in the gas on phase the $\langle T_{Cu}\rangle$ values measured using different tools agree with each other and present similar behavior, that is, both increase as $d$ is incremented. However, in the gas off phase the $T_{gas}$ values measured with the FOT sensor are approximately 3 °C lower than the ones measured using the IR cam. On the other hand, the $P_{th-gas}$ values measured with the IR cam are lower than the ones measured with the FOT sensor, which is probably a consequence of the smaller difference between the $T_{gas}$ values in the gas on and off phases. The $P_{th-gas}$ curves in figure 7 looks to reach a plateau for $d \geqslant 20.0$ mm, but this is more evident for the curve obtained with the IR cam. This plateau in $P_{th-gas}$ may indicate that, starting from a certain point, the contribution from the gas to the generation of thermal power is lost, probably to the ambient air, instead of being transferred to the Cu plate.

Measurements of $\langle T_{Cu}\rangle$ and $\langle T_{gas}\rangle$ values as a function of $d$ were carried out also producing a plasma jet. For this purpose, $T_{Cu}$ and $T_{gas}$ values were recorded, simultaneously, as a function of time positioning the FOT sensors according to the scheme shown in figure 2. From such data, not only the $\langle T_{Cu}\rangle$ values were obtained from the plateau of the plasma on phase, but also the $\langle T_{gas}\rangle$ ones. In addition, the values of thermal output when producing a plasma jet ($P_{th}$) were also extracted from the inclination of the $T_{Cu}$ versus time curve right after the plasma discharge is switched on. Figure 8 shows the results obtained in this experiment. In 8(a) it is shown the temporal evolution of both $T_{Cu}$ and $T_{gas}$ for a distance $d = 20.0$ mm between the plasma outlet and the Cu plate. In 8(b) it is presented the variation of $P_{th}$ as a function of $d$. In 8(c) are presented the $\langle T_{Cu}\rangle$ and $\langle T_{gas}\rangle$ versus $d$ curves when the Cu plate is grounded through the RC-circuit. The same parameters as in 8(c) but with the Cu plate grounded without the RC-circuit are presented in 8(d). Photographs of the He plasma jet impinging on the Cu plate for $d = 10$ mm, 26 mm and 38 mm are shown in figure 9.

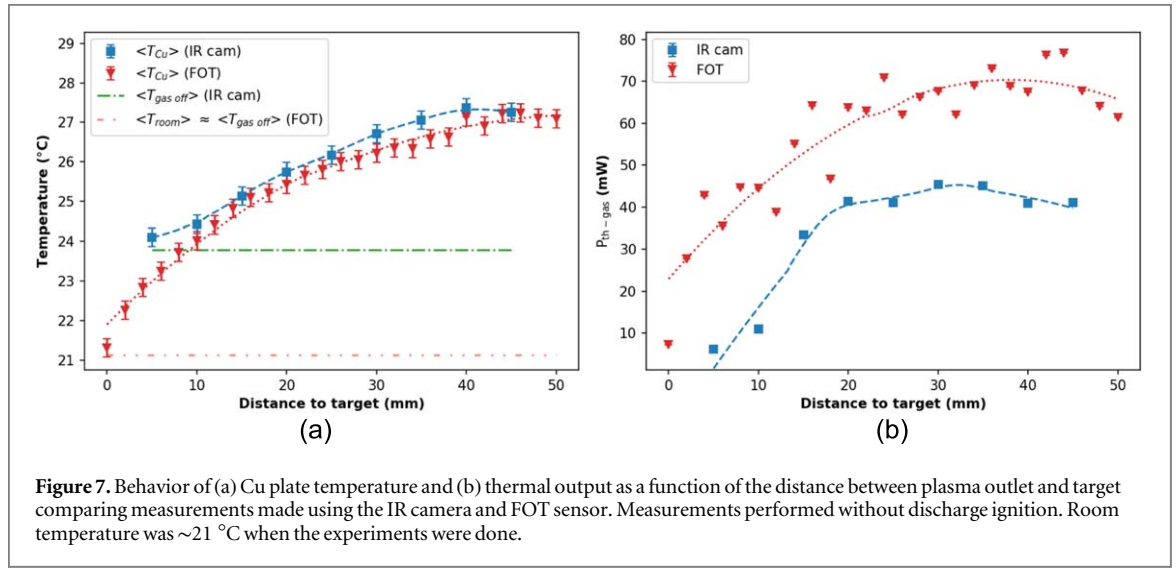

**Figure 7.** Behavior of (a) Cu plate temperature and (b) thermal output as a function of the distance between plasma outlet and target comparing measurements made using the IR camera and FOT sensor. Measurements performed without discharge ignition. Room temperature was ∼21 °C when the experiments were done.

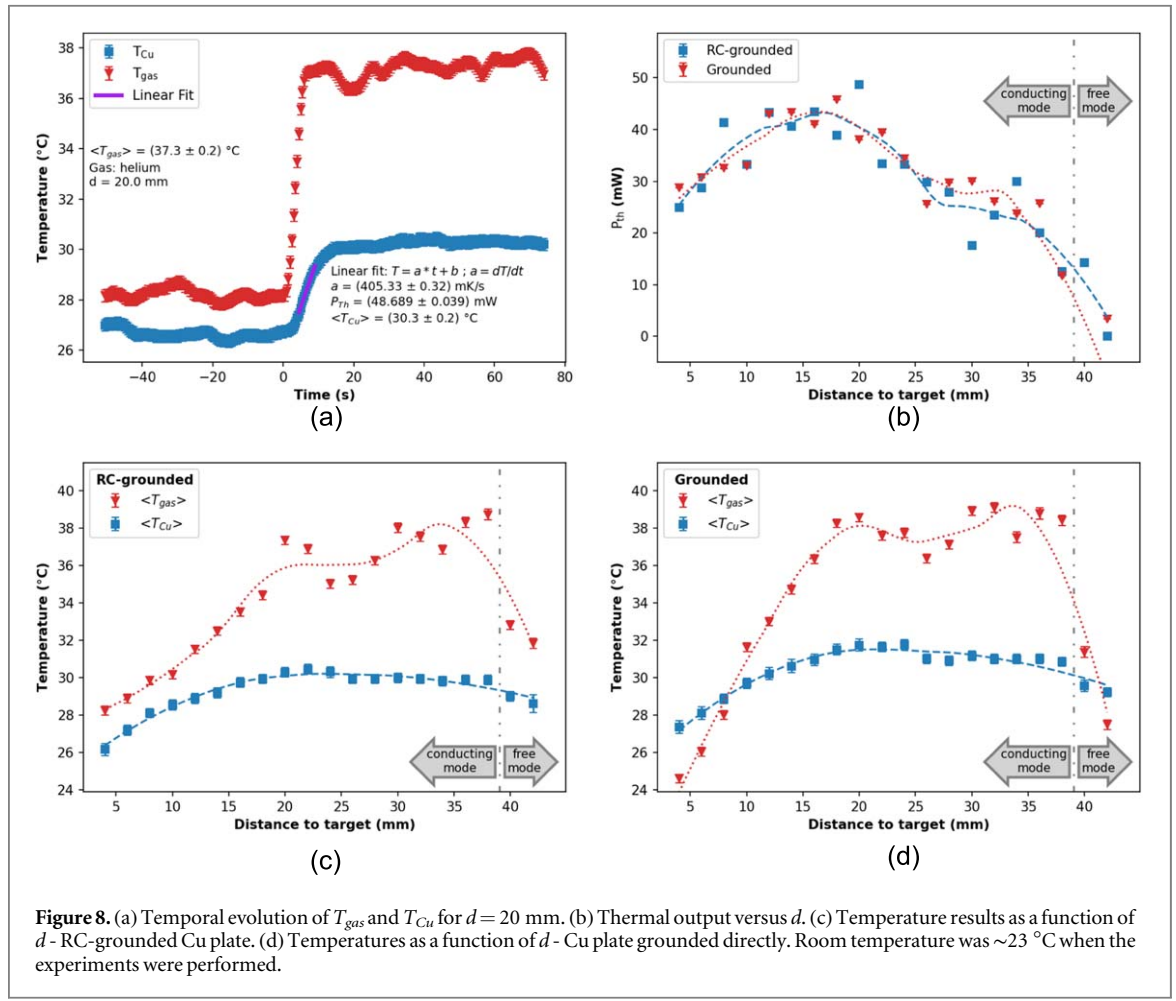

**Figure 8.** (a) Temporal evolution of $T_{gas}$ and $T_{Cu}$ for $d = 20$ mm. (b) Thermal output versus $d$. (c) Temperature results as a function of $d$ - RC-grounded Cu plate. (d) Temperatures as a function of $d$ - Cu plate grounded directly. Room temperature was ∼23 °C when the experiments were performed.

From figure 8(a) it can be seen that when the plasma is switched on, at $t \approx 0$ s, the $T_{gas}$ values increase much more and faster than the $T_{Cu}$ ones. The change in the gas temperature when the plasma is switched on means that the electrical energy used to ignite the plasma is one of the heating sources of the APPJ, as it was expected. From the $P_{th}$ curves in figure 8(b) it can be seen that for $d \leqslant 20.0$ mm $P_{th}$ has a growth trend but for $d > 20.0$ mm it starts to decrease. Figure 8(c and d) shows that the $\langle T_{gas}\rangle$ values are higher than the $\langle T_{Cu}\rangle$ in all the distance range for which the temperatures were measured. From that figure it can be noticed that the $\langle T_{gas}\rangle$ values present a growth trend until a certain $d$ value ($d = 38$ mm, for instance). From this point, the APPJ does a transition from the conducting mode, in which the plasma touches the Cu plate, to the free mode (plasma stops touching the target). Such growth trend is in agreement with the results obtained in the free jet condition. When comparing the $\langle T_{gas} \rangle$ and $\langle T_{Cu} \rangle$ curves in figure 8(c and d) it is observed that while the first present a growth trend, the second looks to reach a plateau for $d > 20.0$ mm. It is an interesting point indicating that even if the APPJ temperature is still growing, at some point part of the thermal energy from the plasma is not transferred to the target anymore and it is probably dissipated in the ambient gas. This is partially confirmed by the results of $P_{th}$ as a function of $d$ shown in figure 8(b).

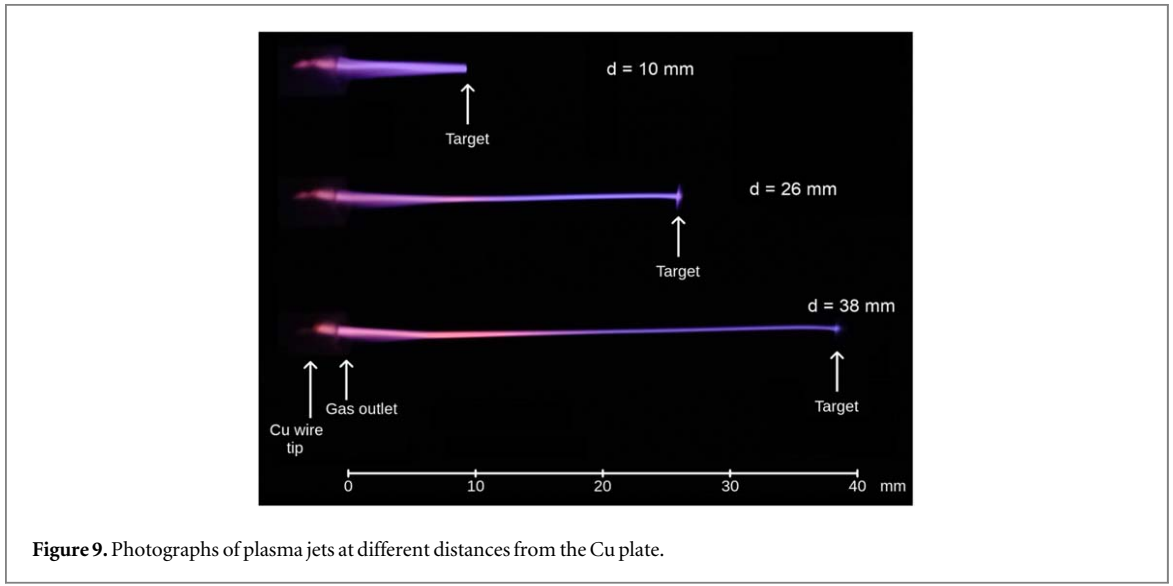

**Figure 9.** Photographs of plasma jets at different distances from the Cu plate.

It is important to mention that the $T_{gas}$ measurements were performed with the FOT sensor positioned so that its tip was always in the center of the plasma column. Thus, due to this reason, it is possible that the centers of the gas and plasma column were not exactly the same.

By comparing the thermal output values obtained without discharge ignition (see figure 7(b)) with the ones achieved producing a plasma jet (figure 8(b)) it can be observed that the $P_{th-gas}$ is noticeably higher than the $P_{th}$ ones. So, an important remark must be made here: when the measurements are performed producing a plasma jet the gas is already on, flowing and impinging on the Cu plate before the plasma is switched on, which makes the $T_{Cu}$ values be higher than $T_{room}$ for $t \leqslant 0$ in the temperatures *vs* time curves. This leads to a smaller variation in the $T_{Cu}$ values when the plasma is ignited and can also affect the inclination of the $T_{Cu}$ versus $t$ curves right after the discharge ignition, resulting in $P_{th}$ values lower than the $P_{th-gas}$ ones. Anyway, this enforces the idea that the main heating source in the He APPJ is the helium gas itself. Of course, other possible factors, like the production of reactive species and vibrational excitation of the $N_2$ molecules, could be affecting the gas temperature in the APPJ [57–59]. However, through OES measurements (not included in this work) we noticed that the amount of reactive species produced within the plasma jet increases only up $d = 12.0$ mm, presenting a reduction trend after that. Measurements of vibrational temperatures ($T_{vib}$) assessed through OES were also obtained as a preliminary investigation of this work. From such measurements it was found that when operating in the free jet mode, the vibrational excitation of $N_2$ molecules do not seems to have any correlation with the behavior of the $T_{gas}$ as a function of $d$. However, when the plasma jet impinges on a conductive target, it was observed that the $T_{vib}$ values measured as a function of $d$ present a behavior very similar to the ones for $T_{gas}$ shown in figure 8(c,d). Thus, we can infer that the vibrational excitation of the $N_2$ molecules may be an additional heating source of the He APPJ when it impinges on a conducting target.

In some previous works that studied the behavior of $T_{gas}$ as a function of $d$, it was observed a reduction in the $T_{gas}$ values as $d$ was incremented [24, 26, 60]. However, in the work by Lotfy *et al* [60] they employed a He flow rate of 14 slm in their experiments. Such $Q$ value is very high and it is probably above the $Q$ range in which the temperature grows when increasing $d$. Nguyen *et al* [26] employed a $Q$ value of 2.0 slm to produce an APPJ in a coaxial double dielectric barrier discharge reactor. In their experiments, a decay in the $T_{gas}$ values from $\sim$85 °C at $d = 5.0$ mm to $\sim$50 °C at $d = 20.0$ mm was observed. In the work by Nastuta and Gerling [24] they employed a He flow rate of 2.0 slm to generate an APPJ using a ring electrode reactor. They observed a clear reduction trend in $T_{gas}$ as $d$ was incremented only when $T_{gas}$ was higher than 120 °C at $d = 0.0$ mm and decreased to nearly 50 °C at $d = 20.0$ mm. When $T_{gas}$ was $\sim$45 °C at $d = 0.0$ mm, the temperature increased to $\sim$52 °C for $1.0 \leqslant d \leqslant 5.0$ mm and, after that, it decreased monotonically until reach a value of $\sim$40 °C at $d = 20.0$ mm. Therefore, in the works by Nguyen *et al* or Nastuta and Gerling the intrinsic He gas warming effect was not perceived probably because of the high $T_{gas}$ achieved when producing a plasma jet with their devices. A common feature of the plasma sources employed in those three works was the use of power supplies operating at frequencies higher than 15 kHz. Being that, both Nguyen and Nastuta&Gerling reported discharge power of the order of 1.0 W.

Kim *et al* [41] measured $T_{gas}$ as a function of *d* for a plasma candle using a FOT sensor. In their work they produced a He plasma jet using a flow rate of 10 slm. The authors reported an almost constant temperature profile, close to 33 °C, as a function of *d*. Here again, the high gas flow rate may have been the reason why the heating of the plasma jet intrinsic to the He gas was not noticed.

Moon and Choe [33], using He as the working gas, observed an increment in the gas temperature at different axial positions in a plasma jet impinging on a grounded electrode. The $T_{gas}$ measurements were performed employing both OES and a thermocouple, for a fixed *d* value but at different axial positions. In their work, Moon and Choe found a linear temperature increment, from ~780 K to ~930 K in the range $0.0 < d < 20.0$ mm. On that occasion, the authors attributed the increment in the $T_{gas}$ values to the emission of secondary electrons by the grounded electrode.

In one of our previous works within the group, a PhD thesis by Nishime [61], He was employed as the carrier gas at a flow rate of 1.5 slm to produce a plasma jet in free mode. The $T_{gas}$ measurements were performed using a FOT sensor and a significant increment in the $T_{gas}$ values as *d* was increased was observed. In fact, the $T_{gas}$ vs *d* curves obtained by Nishime (see figure 4.21 in [61]) presented a trend very similar to the one measured in this work for the plasma jet in free mode (see figure 3(a)). In that work $T_{gas}$ was not measured with the plasma switched off. Thus, the explanation for the increase in the $T_{gas}$ values for large *d* in the occasion was that most ionization processes occur at the tip of the plasma plume due do a higher density of He atoms in metastable states. So, an increment in $T_{gas}$ after the end of the plasma plume would be reasonable. However, as it has been shown in figure 3(a), the $T_{gas}$ curves as a function of *d* measured with and without discharge ignition presented similar values and behavior. Based on this we can conclude that in the free jet case and using the mentioned plasma source, there are no significant influence of ionization processes on the increment of the gas temperature values as *d* is increased.

Measurements of $T_{gas}$ in He APPJ devices with plasma on and off and as a function of the distance from the gas outlet were carried out by Slikboer *et al* [39] and also by Hofmans in a PhD thesis [62]. In both works the authors used a dielectric capillary with the same geometry (outer and inner diameters equal to 4.0mm and 2.5 mm, respectively) and a FOT sensor to build a two-dimensional (2D) map off the gas temperature in the gas/plasma flow and in its vicinity. The works differ in the gas flow rate and flow direction. Slikboer *et al* placed the capillary tube horizontally and employed a 0.7 slm He flow, while Hofmans positioned the tube vertically with the He gas flow pointing downwards at a flow rate of 1.5 slm. Slikboer *et al* observed an increment in the $T_{gas}$ values in the core of the gas stream as *d* was increased for both plasma on and off cases. The $T_{gas}$ measurements presented in their 2D temperature map were limited to a range of ~14 mm. The measurements performed by Hofmans has a longer range in the axial direction (45.0 mm). In that PhD thesis it was also observed an increment in the $T_{gas}$ values as *d* was increased, for both plasma on and off cases. However, in that work $T_{gas}$ increased only up to $d \approx 10$ mm with plasma off and up to $d \approx 12$ mm with plasma on, and started to decrease after that. This last result is likely a consequence of the gas flow pointing downwards.

Thus, even with the increment in $T_{gas}$ as a function of the distance to device outlet having been observed in previous studies, the fact that part of the gas heating in APPJs comes just from the He gas itself has not been systematically investigated before. Therefore, the present work is the first one to provide a more detailed investigation of the phenomenon in the context of generation of atmospheric pressure plasma jets.

### 3.2.1. Electrical parameters for He plasma discharge

In figure 10 are shown main electrical parameters of the discharge. In (a) there is an example of the voltage and current waveforms measured for the APPJ when the distance between the plasma outlet and target was 20.0 mm, and in (b) are presented the voltage amplitude ($V_{pp}$), discharge power ($P_{dis}$) and effective current ($i_{RMS}$), averaged over 10 consecutive measurements, as a function of the distance from target. Those measurements were performed for the case in which the Cu plate was grounded through the RC-circuit. The $P_{dis}$ values obtained for *d* values lower than 25.0 mm are of the order of 1.0 W. In figure 10(b) a weak growth trend in the $P_{dis}$ values is noticed for $d < 15.0$ mm. Such growth trend is followed by a clear downward trend after that, until a transition from the conducting to the free mode, when the plasma jet stops touching the Cu plate, is achieved for $d > 38.0$ mm. The $i_{RMS}$ values, in turn, present an almost monotonically decreasing behavior as *d* is incremented, from ~5.5 mA to ~4.0 mA, until the discharge changes to the free mode, when there is an abrupt reduction in the $i_{RMS}$ values.

As it can be seen in figure 10(b), the $P_{dis}$ and $i_{RMS}$ curves behave slightly different. This happens because the power supply is sensitive to impedance matching, which causes a drop in the amplitude of the applied voltage. The magnitude of the voltage drop is considerably higher at small *d* values, in a way that the discharges do not ignite properly for $d < 4.0$ mm. Thus, the growth in the $P_{dis}$ values for $d$ between 4.0 mm and 14.0 mm is mainly due to a reduction in the voltage drop, which makes the applied voltage be higher as $d$ increases.

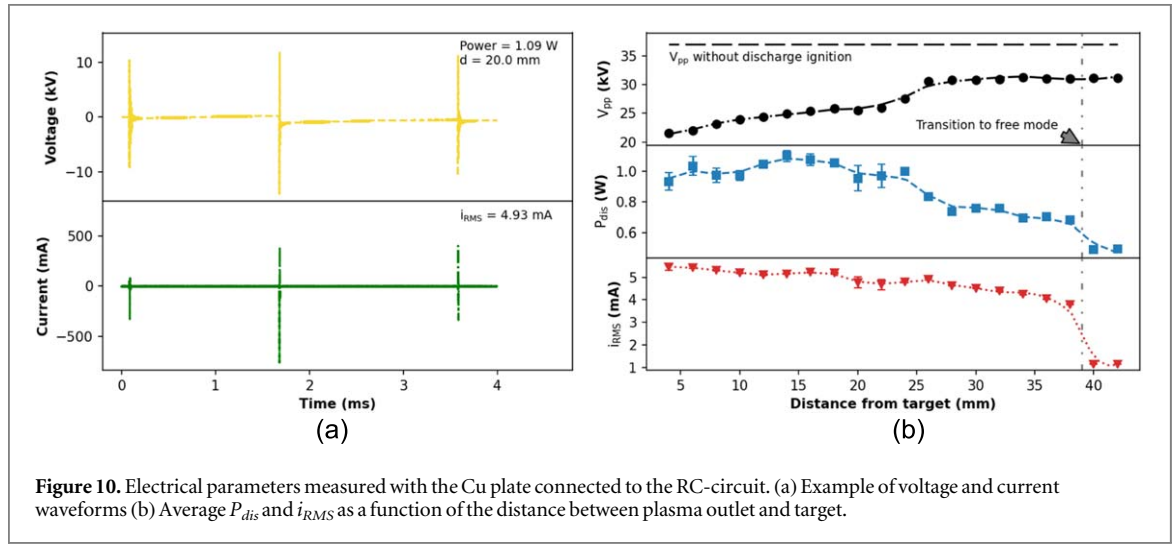

**Figure 10.** Electrical parameters measured with the Cu plate connected to the RC-circuit. (a) Example of voltage and current waveforms (b) Average $P_{dis}$ and $i_{RMS}$ as a function of the distance between plasma outlet and target.

If we compare the $P_{th}$ and $P_{dis}$ curves as a function of $d$ (figures 8(b) and 10(b), respectively), it can be observed a relationship between the thermal output and the discharge power, since the behavior of both curves are similar. That is, both present a growth trend at the beginning followed by a downward in the values as $d$ is increased. However, the peak $P_{dis}$ value occurs at a distance $d$ lower than the one where the peak of $P_{th}$ is observed. If we look at the $P_{th-gas}$ curve in figure 7(b), we see that the contribution of the thermal output intrinsic to the He gas is still growing up for $d > 20.0$ mm. Thus, the 'retarded' peak in the $P_{th}$ versus $d$ curve is probably due to the contribution of the $P_{th-gas}$ for larger $d$ values.

A point to be highlighted regarding the influence of the discharge power on the $T_{gas}$ values is that despite the fact of $P_{dis}$ in this work be of the same order of the values reported by Nguyen and Nastuta&Gerling, that is, $P_{dis} \approx 1.0$ W, $T_{gas}$ was always lower than 40 °C [24, 26]. This last result can be attributed mainly to the low repetition rate of the voltage pulses of the power supply used in this work, 50Hz, which is much lower than the ones used in the mentioned works (>15 kHz).

## 4. Conclusions

In this work, measurements of gas temperature were performed using He as the working gas in the free gas/jet mode as well as with the gas/plasma impinging on a Cu surface. Some measurements were also performed using Ar and $CO_2$ as working gases, but with discharge ignition only for Ar. Temperature was investigated via fiber optics thermometer for gas and surface temperature as well as with infrared camera measurement of the surface temperature.

When using He as the carrier gas, from the gas temperature measurements performed with and without producing a gas discharge in the free mode, it was found that the $T_{gas}$ values changes as a function of the distance from the outlet, most times growing up as $d$ increases independent whether a discharge was ignited or not. In some cases the temperature increment was of the order of 10 °C. Such temperature variation was not observed when the $T_{gas}$ measurements were performed for Ar or $CO_2$. This leads to a main conclusion that the main heating source of a cold He-APPJ can be the helium gas expansion itself. In addition, the proportion of increment in $T_{gas}$ depends mainly on the distance from the gas/plasma outlet.

When in the free mode, it was observed that the gas temperature does not change significantly between the plasma on and off cases with He as the carrier gas, and when using Ar the gas temperature value is very close to $T_{room}$. This suggests that the device employed in this work is not able to heat up the plasma jet in free mode. On the other hand, when the He-APPJ impinges on the Cu target there is a significant change in both gas and target temperatures when comparing the plasma off and on states. The comparison between the free and conductive jet conditions, for a fixed distance from the gas outlet, indicates that the other important heating sources of the gas discharge are related to electrical parameters that change when the APPJ hits a conducting surface. In this case, when the plasma touches the target there is an electrical current passing through it, dissipating energy in the discharge which results in additional gas warming.

Through the comparison of the plasma off and on cases, with the gas/plasma impinging on the Cu plate, it was possible to verify that in both cases the gas also transfers heat to the Cu plate in both plasma off and on conditions. However, from a certain distance (nearly 20.0 mm in our experiments) and with discharge ignition, the increase in plasma temperature does not result in an increase in the temperature of the target. Nevertheless, when the measurements are performed just with the He flow (without discharge ignition) the Cu plate temperature keeps growing up for $d > 20.0$ mm, but at a smaller rate.

Based on the $T_{gas}$ data obtained in the experiments, we can infer, at first, that the gas heating with the increment in the distance from outlet is influenced by both Joule-Thomson, buoyancy and Dufour effects. However, further investigation performing temperature measurements simultaneously with Schlieren imaging would provide more precise results and, possibly, additional data to explain the temperature values growth as a function of the distance from the gas outlet when using He.

## Acknowledgments

The authors thank to Bruno Honnorat (INP-Greifswald) for the fruitful discussions that helped us to improve this work. This work was supported by the São Paulo Research Foundation-FAPESP (grants # 2019/05856-7 and # 2021/14391-8).

## Data availability statement

All data that support the findings of this study are included within the article (and any supplementary files).

## ORCID iDs

Fellype do Nascimento  https://orcid.org/0000-0002-8641-9894
Torsten Gerling  https://orcid.org/0000-0002-5184-257X
Konstantin Georgiev Kostov  https://orcid.org/0000-0002-9821-8088